\documentclass[a4paper]{pos}

\usepackage{subfigure}

\newcommand{\Cgrav}{\ensuremath{C_{\rm grav}}}

\title{Expected Performance of the ATLAS Detector in GMSB Models with
  Tau Final States}

\ShortTitle{GMSB models with $\tau$ final states}

\author{\speaker{D\"orthe Ludwig} on behalf of the ATLAS collaboration\\
  DESY Hamburg/ University of Hamburg, Germany\\
  E-mail: \email{dorthe.ludwig@cern.ch}}

\abstract{Gauge Mediated Supersymmetry Breaking (GMSB) models provide
  a possible mechanism to mediate Supersymmetry to the visible sector.
  In these models the lightest supersymmetric particle (LSP) is
  usually the gravitino, while the next-to-lightest supersymmetric
  particle (NLSP) is either a neutralino or a slepton. In the case of
  a stau NLSP events with large missing transverse energy, highly
  energetic jets and up to four $\tau$ leptons are expected in
  $pp$-collisions at the LHC. A study of the expected performance of
  the ATLAS detector in GMSB scenarios with a stau NLSP for a LHC
  center-of-mass energy of $\sqrt{s}$ = 10\,TeV is presented. A
  cut-based selection has been optimised using a typical GMSB scenario
  and a scan of the GMSB parameter space has been performed to
  determine the discovery reach as a function of the integrated
  luminosity. In addition, the invariant mass distribution of two
  $\tau$ leptons has been used to study the measurement of masses of
  supersymmetric particles with larger event samples.}

\FullConference{XXth Hadron Collider Physics Symposium \\
		 November 16 -- 20, 2009 \\
		 Evian, France}

\begin{document}

\section{GMSB Models}

In GMSB models Supersymmetry is communicated from the secluded sector
to the visible sector through a flavor-blind SM gauge interaction via
messenger fields at a scale $M_m$ small compared to the Planck mass.
Squarks, sleptons and gauginos obtain their masses radiatively from
the gauge interactions with the massive messenger fields. The minimal
GMSB model ist characterised by six fundamental parameters: the
effective SUSY breaking scale $\Lambda$, the mass of the messengers
$M_m$, the number of messenger SU(5) supermultiplets $N_5$, the ratio
of the Higgs vacuum expectation values tan$\beta$, the sign of the
Higgsino mass term sgn$\mu$ and the scale factor of the gravitino
coupling \Cgrav which determines the NLSP lifetime. If R-parity is
assumed to be conserved the lightest supersymmetric particle (LSP),
the gravitino $\widetilde{G}$, is stable. In large parts of the
parameter space the next-to-lightest supersymmetric particle (NLSP) is
the $\tilde{\tau}$.  An example is the scenario called GMSB6 with
$\Lambda$=40\,TeV, $M_m$=250\,TeV, $N_5$=3, $\tan\beta$=30, $\rm
sgn\mu$=+ and \Cgrav=1, where $\rm m_{\tilde{\tau}_1}$=102.8\,GeV.
GMSB models with $\tilde{\tau}_1$ NLSP have been searched for at
LEP~\cite{bib:LEPexl}. For prompt decays, $\tilde{\tau}_1$ NLSPs with
masses below 87\,GeV have been excluded.

\section{Study of the Discovery Potential in the GMSB Parameter Space}

In regions of the GMSB parameter space where the NLSP is the
$\tilde{\tau}_1$, long cascade decays of the initial squarks and
gluinos lead to many highly energetic jets, many $\tau$ leptons, and a
significant amount of missing transverse energy ($E_T^{\rm miss}$) due
to the escaping $\widetilde{G}$. For this reason the following
preselection is applied: events must pass the trigger selection
containing at least one jet with $p_T >$ 70\,GeV and $E_T^{\rm miss}
>$ 30\,GeV, two or more reconstructed jets have to be found (leading
jet $p_T >$ 100\,GeV, second-leading jet $p_T >$ 50\,GeV), at least
one hadronically decaying $\tau$ lepton has to be found (leading $\tau\,
p_T >$ 20\,GeV), $E_T^{\rm miss}$ has to exceed 60\,GeV and the
azimuthal angle between the leading jet and the direction of $E_T^{\rm
  miss}$ needs to exceed 0.2. This selection yields a signal
efficiency of 43\% for the GMSB6 scenario. The remaining SM background
mainly consists of $t\bar{t}$ and $W$ events.  For further suppression
of this SM background a two-dimensional optimisation of
$S=N_S/\sqrt{N_B}$, where $N_S$ $(N_B)$ is the number of signal
(background) events, has been performed. The maximum significance $S$
can be achieved for $E_T^{\rm miss} >$ 280\,GeV and $N_{\tau}\ge 2$
yielding $20.4\pm0.7$ signal events for the GMSB6 scenario and
$2.5\pm1.5$ expected background events for $\rm {\cal
  L}=200\,pb^{-1}$~\cite{bib:GMSBpubnote}.

\begin{figure}
 \centering
 \subfigure[]{%
 \label{fig:scan1}%
 \includegraphics[width=0.49\textwidth]{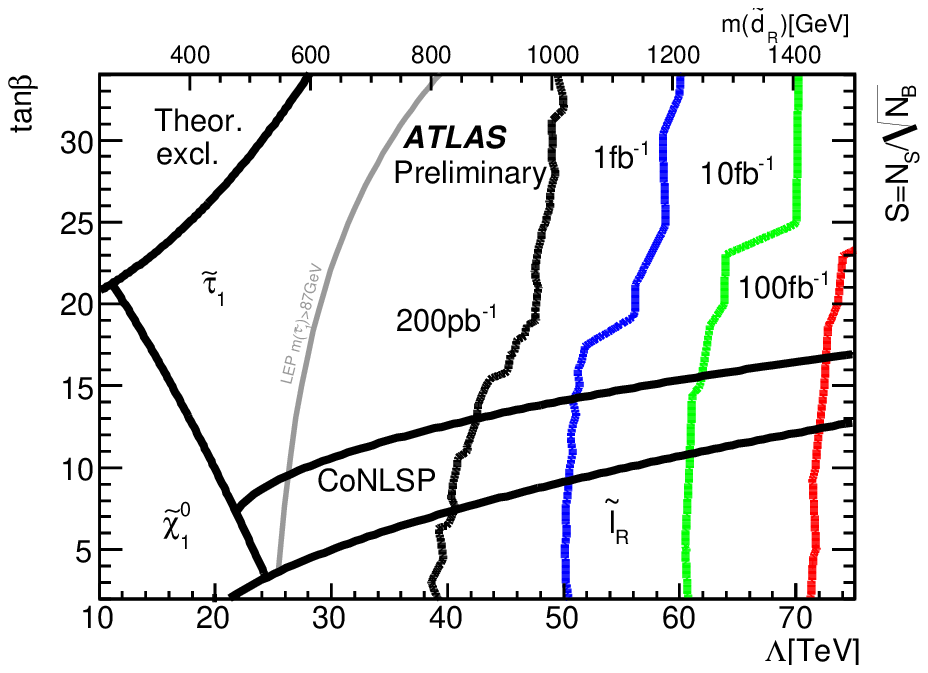}}
 \subfigure[]{%
 \label{fig:scan2}%
 \includegraphics[width=0.49\textwidth]{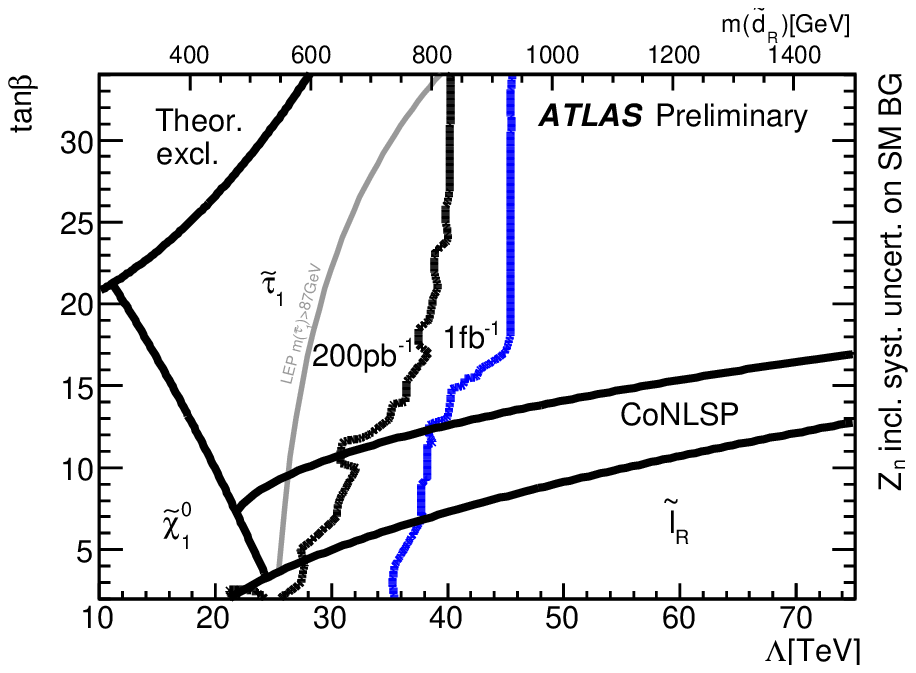}}
\caption{Integrated luminosity needed for a signal significance of
  $S=5$ or $Z_n=5$, respectively, in the ($\Lambda$-$\tan\beta$)-plane
  for $M_m=250\,TeV$, $N_5=3$, $\rm sgn\mu=+$ and $\Cgrav=1$ using (a)
  the simple calculation of the significance $S$ and (b) $Z_n$
  properly including the systematic uncertainties.}
 \label{fig:ScanGMSB}
\end{figure}

The discovery potential in the GMSB parameter space is studied in the
($\Lambda$-$\tan\beta$)-plane for $M_m=250$\,TeV, $N_5=3$, \mbox{$\rm
  sgn\mu=+$} and $\Cgrav=1$. These parameter values restrict the
analysis to specific, promptly decaying NLSPs. The number of selected
signal events in the ($\Lambda-\tan\beta$)-plane for $\rm
200\,pb^{-1}$ and the expected number of background events can be
translated into a signal significance as a function of the integrated
luminosity $\cal L$. The corresponding results are shown in
Fig.~\ref{fig:scan1}. However, this simple definition neglects the
influence of systematic uncertainties on the background expectation.
These relative uncertainties have been estimated to be 50\% for a
centre-of-mass energy of 10\,TeV. A more appropriate calculation of
the significance $Z_n$ including these systematic
uncertainties~\cite{bib:CSCBook} (p.1590) provides a more conservative
estimate of the signal significance as displayed in
Fig.~\ref{fig:scan2}. This significance definition reduces the
parameter region for a 5$\sigma$ discovery with $\rm {\cal
  L}=200\,pb^{-1}$ ($\rm 1\,fb^{-1}$) from $\Lambda\sim$ 50\,TeV
(60\,TeV) to $\Lambda\sim$ 40\,TeV (45\,TeV). The discovery reach is
limited by the systematic uncertainty of the background.

\section{Study of the Invariant Di-Tau Mass Distribution}

After a possible SUSY discovery the determination of the masses of
SUSY particles is vital. In the presence of two undetected LSP only
kinematic end points of invariant mass distibutions can be measured,
mostly sensitive to differences of SUSY masses. For the study of the
kinematic end-point of the invariant mass of two $\tau$ leptons larger
datasets ($\rm 8\,fb^{-1}$) are considered. The selection is slightly
loosened compared to the one mentioned in the previous section to
allow for sufficient statistics~\cite{bib:GMSBpubnote}. Due to the
unmeasured neutrinos in $\tau$ decays the characteristic triangular
shape and therefore the kinematic end-point of the di-tau invariant
mass distribution is lost impeding its direct extraction.  For this
reason the combined OS-SS distribution of the GMSB signal and the SM
background is fitted to extract the inflection point $m^{\rm
  IP}_{\tau\tau} $ of the distribution which is translated into the
end-point $m_{\tau\tau}^{\rm max}$ using a linear calibration curve
following the method proposed in~\cite{bib:CSCBook}~(p.~1617). The
determination of the end-point is subject to several sources of
systematic uncertainty. Combining all systematic unceratinties the
following kinematic endpoint is obtained:
\begin{equation}
  m_{\tau\tau}^{\rm max}  = \left(\,135\pm 4 \,({\rm stat.})\,^{+13}_{-\,9}\,({\rm sys.})\,\pm13\,({\rm SUSY\,model})\,\right) \rm \,GeV\,.
\end{equation}
This result demonstrates that a measurement of the end-point of the
invariant di-tau mass spectrum might be possible in the GMSB6 scenario
with a small bias from additional SUSY background.

\end{document}